\documentclass[twocolumn]{revtex4-1}
\usepackage{graphicx} 
\usepackage[utf8]{inputenc}
\usepackage[english]{babel}
\usepackage[T1]{fontenc}
\usepackage{amsmath}
\usepackage{float}
\usepackage{hyperref}
\usepackage{lipsum}
\usepackage{natbib}

\begin{document}

\title{Impact of Decoherence on Average Correlation}

\author{A. M. Silva}
\affiliation{Instituto de Física Teórica, Universidade Estadual Paulista, Rua Dr. Bento Teobaldo Ferraz, 271, Bloco II, São Paulo, 01140-070, SP, Brazil}

\begin{abstract}
This article presents a comprehensive study of the impact of decoherence on the average correlation for pure quantum states. We explore two primary mechanisms of decoherence: phase damping and amplitude damping, each having distinct effects on quantum systems. Phase damping, which describes the loss of quantum coherence without energy loss, primarily affects the phase relationships between the components of a quantum system while amplitude damping involves energy dissipation and also affects the state's occupation probabilities. We show that the average correlation follows a predictable decaying pattern in both scenarios. Our analysis can be understood in the context of quantum computing, by focusing on how phase damping influences the entanglement and correlation between qubits, key factors in quantum computational efficiency and error correction protocols.

\end{abstract}

\maketitle

\section{Introduction}

The study of quantum mechanics and its applications in quantum computing and quantum information theory has highlighted the pivotal role of coherence in quantum systems. Decoherence, the process by which quantum systems interact with their environment leading to the loss of quantum coherence, poses significant challenges in maintaining the integrity of quantum states \cite{decoherence1,decoherence2,decoherence3}.
Additionally, measures of coherence and nonclassicality are critical in understanding the transition from quantum to classical behavior in physical systems \cite{measure1,measure2,measure3,measure4}. Coherence, indicative of the superposition states in quantum mechanics, is a key feature that differentiates quantum systems from classical ones. It allows for phenomena like interference, which are absent in classical systems. Nonclassicality on the other hand extends beyond coherence, encompassing aspects like quantum entanglement and violations of Bell inequalities, which defy classical explanations based on local realism and deterministic probabilities \cite{nonclassical1,nonclassical2}.

Traditionally, nonclassicality has been quantified using Bell inequalities, which compare quantum mechanical predictions against classical physics predictions for certain physical systems. However, this approach, while effective, often demands precise control over experimental parameters, posing practical challenges in laboratory settings \cite{labs}.

A recent approach suggested the use of average correlation as an alternative measure \cite{Tschaffon2023}. This innovative approach does not necessitate stringent control over measurement directions and offers ease of computation, making it highly suitable for practical applications. The paper meticulously articulates the derivation of average correlation for various quantum states, including pure states and Werner states, providing a thorough understanding of how these states exhibit nonclassicality under specific conditions.


Moreover, the study delves into the relationship between Bell inequalities and correlation functions of dichotomic observables, laying a solid theoretical foundation for the proposed methodology. The authors' analysis includes an extensive mapping of two-qubit states based on average correlation and the Bell parameter. This mapping is instrumental in distinguishing between classical and nonclassical states and offers profound insights into the behavior of mixed states in quantum systems \cite{Tschaffon2023}. Within this context we aim to analyze the impact of decoherence on average correlation, by initially focusing on pure quantum states.

Decoherence is not only a central theme in the study of quantum mechanics but also a critical barrier in the development of quantum computing and information processing technologies \cite{tech1,tech2,tech3,tech4,tech5}. Our study aims to examine how decoherence affects average correlation, thereby influencing the nonclassical characteristics of quantum states. This analysis is particularly relevant in the context of quantum computing and information processing, where maintaining nonclassicality is essential for harnessing the full potential of quantum technologies.

\section{Average Correlation for Pure States}
Starting with the Schmidt decomposition for pure quantum states:
\begin{equation}
    |\psi\rangle =c|0\rangle_A |1\rangle_B - \sqrt{1-c^2}|1\rangle_A |0\rangle_B
\end{equation}
We wish to evaluate the average correlation given by:
\begin{equation}
    \Sigma=\frac{1}{4\pi^2}\int d\Omega_\textbf{a} \int d\Omega_\textbf{b} |E(\textbf{a},\textbf{b})|
\end{equation}
where $E(\textbf{a},\textbf{b})$ denotes the correlation function measured by two observers A and B and \textbf{a} and \textbf{b} are unit vectors which denote the axis in which the measurement of each observer is made:
\begin{equation}
    E(a,b)=\textbf{a}^T K \textbf{b}
\end{equation}
the correlation matrix $K$ in the formula above is defined as:
\begin{equation}
    K_{ij}=\text{tr}(\hat{\rho}\;\hat{\sigma}_{iA}\otimes \hat{\sigma}_{jB})
\end{equation}
with the corresponding Pauli matrices for each axis direction given by
$\hat{\sigma}_A=(\hat{\sigma}_{1A},\hat{\sigma}_{2A},\hat{\sigma}_{3A})$ and $\hat{\sigma}_B=(\hat{\sigma}_{1B},\hat{\sigma}_{2B},\hat{\sigma}_{3B})$.
For a pure state of the form at Eq. (1), the density matrix can be evaluated and shown to equal:
\begin{equation}
\rho = \begin{pmatrix}
0 & 0 & 0 & 0 \\
0 & c^2 & -c\sqrt{1 - c^2} & 0 \\
0 & -c\sqrt{1 - c^2} & 1 - c^2 & 0 \\
0 & 0 & 0 & 0
\end{pmatrix}
\end{equation}
This leads to a correlation matrix given by:
\begin{equation}
K = \begin{pmatrix}
-2c\sqrt{1-c^2} & 0 & 0 \\
0 & -2c\sqrt{1-c^2} & 0 \\
0 & 0 & -1
\end{pmatrix}
\end{equation}
It is clear that there exits a singular value decomposition of the form:
\begin{equation}
K = 
\begin{pmatrix}
0 & 0 & -1 \\
0 & -1 & 0 \\
-1 & 0 & 0
\end{pmatrix}
\begin{pmatrix}
\alpha & 0 & 0 \\
0 & \beta & 0 \\
0 & 0 & \gamma
\end{pmatrix}
\begin{pmatrix}
0 & 0 & 1 \\
0 & 1 & 0 \\
1 & 0 & 0
\end{pmatrix}
\end{equation}
where $\alpha=1$, and $\beta=\gamma=2c\sqrt{1-c^2}$. Therefore defining:
\begin{equation}
    \tilde{\textbf{a}}=U^T \textbf{a}, \; \tilde{\textbf{b}}=V^T\textbf{b}
\end{equation}
allows us to write the average correlation as:
\begin{equation}
    \Sigma = \frac{1}{16\pi^2}\int d\Omega_\textbf{a} \int d\Omega_\textbf{b} |(\tilde{\textbf{a}})^T \kappa \;\tilde{\textbf{b}}|
\end{equation}
and by redefining $\tilde{\textbf{b}}$ as $\tilde{\textbf{b}}_\kappa =\kappa\tilde{\textbf{b}}$, we may write the scalar product in terms of the length of the vectors and the angle between them such that:
\begin{equation}
\Sigma = \frac{1}{16 \pi^2}\int d\Omega_\textbf{a} \int d\Omega_\textbf{a} |\tilde{\textbf{b}}_\kappa||\cos\theta_{ÀB}|
\end{equation}
If we now choose the $z$-axis of the coordinate system for vector $\tilde{\mathbf{a}}$ to be aligned with that of vector $\tilde{\mathbf{b}}$, the integral becomes independent of the azimuthal angle and we reach the simple expression below:
\begin{equation}
\Sigma = \frac{1}{8\pi} \int d\Omega_{\mathbf{b}}|\tilde{\mathbf{b}}_\kappa|
\end{equation}
Next, introducing spherical coordinates:
\begin{equation}
\tilde{\mathbf{b}}_\kappa =(\gamma \cos\phi \sin\theta, \beta \sin\phi \sin\theta,\alpha \cos \theta)^T
\end{equation}
and the function:
\begin{equation}
f(\phi)=\left(\frac{\beta}{\alpha}\right)^2\sin^2 \phi+\left(\frac{\gamma}{\alpha}\right)^2\cos^2 \phi
\end{equation}
we may write:
\begin{equation}
\Sigma = \frac{\alpha}{8\pi}\int_0^{2\pi}d\phi \int_0^\pi d\theta \sin\theta \sqrt{f(\phi)\sin^2\theta +\cos^2\theta}
\end{equation}
Performing the variable substitution $u=\cos\theta$ leads to:
\begin{equation}
\Sigma =\frac{\alpha}{8\pi}\int_0^{2\pi} d\phi\sqrt{f(\phi)}\int_{-1}^1 du\sqrt{1+\frac{1-f(\phi)}{f(\phi)}u^2}
\end{equation}
and solving the integral over $u$ yields:
\begin{widetext}
\begin{equation}
\Sigma = \frac{\alpha}{4} \left[ 1 + \frac{1}{2\pi} \int_{0}^{2\pi} d\phi \frac{f(\phi)}{\sqrt{1 - f(\phi)}}\; \text{Arsinh} \left( \sqrt{\frac{1 - f(\phi)}{f(\phi)}} \right) \right]    
\end{equation}
\end{widetext}
Since for a pure state $f(\phi)=\beta^2/\alpha^2$, the integral over $\phi$ can be performed immediately leading to:
\begin{equation}
\Sigma = \frac{\alpha}{4} \left[ 1 + \frac{1}{\alpha} \frac{\beta^2}{\sqrt{\alpha^2 - \beta^2}} \;\text{Arsinh} \left( \sqrt{\frac{\alpha^2-\beta^2}{\beta^2}} \right) \right]    
\end{equation}
As shown at \cite{Tschaffon2023} the average correlation in this case has a maximum of $\Sigma =1/2$ for $c=1/\sqrt{2}$ and a minimum of $\Sigma=1/4$ for $c=1$. The authors also show that an average correlation value of \(\Sigma \leq \frac{1}{4}\), is an indication of classical states, whereas \(\Sigma > \frac{1}{2\sqrt{2}}\) is associated exclusively 
with nonclassical states.

\section{Impact of Decoherence}
\subsection{Phase Damping}
We start our analysis by considering the action of the following Kraus operators on the density matrix \cite{Nielsen2012}:
\begin{equation}
K_0 = \begin{pmatrix}
1 & 0 \\
0 & \sqrt{1 - p}
\end{pmatrix}
\quad
K_1 = \begin{pmatrix}
0 & 0 \\
0 & \sqrt{p}
\end{pmatrix}
\end{equation}
Here, \(p\) represents the probability of phase damping occurring within the quantum system. When applying this to the density matrix for the two quantum states we have:
\begin{equation}
\rho'=\sum_{i,j} (K_{i} \otimes K_{j}) \; \hat{\rho} \; (K_{i} \otimes K_{j})
\end{equation}
Leading to:
\begin{equation}
\rho' = \begin{pmatrix}
0 & 0 & 0 & 0 \\
0 & c^2 & -c\sqrt{1 - c^2}(1-p) & 0 \\
0 & -c\sqrt{1 - c^2}(1-p) & 1 - c^2 & 0 \\
0 & 0 & 0 & 0
\end{pmatrix}
\end{equation}
Evaluating the correlation matrix this time leads to:
\begin{equation}
K = \begin{pmatrix}
-2c\sqrt{1-c^2}(1-p) & 0 & 0 \\
0 & -2c\sqrt{1-c^2}(1-p) & 0 \\
0 & 0 & -1
\end{pmatrix}
\end{equation}
This matrix still posses a singular value decomposition, but with a different parameter $\beta = \gamma = 2c\sqrt{1-c^2}(1-p)$. The function $f(\phi)$ will therefore be given by:
\begin{equation}
    f(\phi)=\frac{\beta^2}{\alpha^2} =4c^2(1-c^2)(1-p)^2 
\end{equation}
And knowing that $p = 1-e^{\Gamma t}$, where $\Gamma$ denotes the decoherence rate, we may substitute $f(\phi)$ on equation (16) for the average correlation and analyze how it varies with time. The result is displayed below:
\begin{figure}[H]
    \centering
    \includegraphics[scale=0.49]{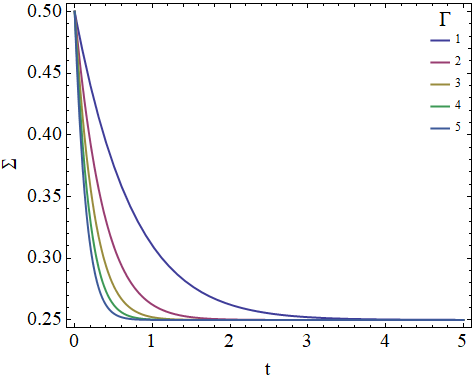}
    \caption{Decay of average correlation over time due to phase damping for various decoherence rates, illustrating that higher rates lead to faster decay in correlation.}
    \label{fig:enter-label}
\end{figure}
We choose to set $c=1/\sqrt{2}$ such that the average correlation starts at the maximum value of $\Sigma =1/2$ and decays with time to the minimum value $\Sigma=1/4$. It is evident that varying the decoherence rates, represented by different values of $\Gamma$, influences the rate at which average correlation diminishes. Higher values of decoherence rate lead to a faster decay of the average correlation as expected. The plot's curves do not fall below $1/4$, which aligns with the observation that pure states, even when entangled, remain nonclassical and do not transition into classical states as they maintain an average correlation above the $1/4$ threshold. This suggests that even as the system experiences phase damping over time, it remains within the regime of nonclassical states.

The analysis reveals that the decay of $\Sigma$ halts at the threshold value of $1/4$, irrespective of the decoherence rate applied. This observation is significant as it indicates that, despite the environmental perturbations modeled by different values of $\Gamma$, the system retains a minimum level of quantum correlation. This behavior underscores the inherent resilience of quantum states to decoherence, suggesting that entangled states, even under adverse conditions, maintain a degree of nonclassicality.

Furthermore, the consistent maintenance of $\Sigma$ above the $1/4$ threshold across different decoherence scenarios posits that the system, subjected to phase damping, persistently resides within the nonclassical regime. This outcome is particularly relevant for quantum information processing, where the preservation of quantum correlations amidst environmental decoherence is crucial for the functionality of quantum technologies.
\subsection{Amplitude Damping}
In the case of amplitude damping, the Kraus operators are given by \cite{Nielsen2012}:
\begin{equation}
K_0 = \begin{pmatrix}
1 & 0 \\
0 & \sqrt{1 - p}
\end{pmatrix}
\quad
K_1 = \begin{pmatrix}
0 & \sqrt{p} \\
0 & 0
\end{pmatrix}
\end{equation}
We may apply these operators to the density matrix for pure state as done in Eq. (19) leading to:
\begin{equation}
\rho' = \begin{pmatrix}
p & 0 & 0 & 0 \\
0 & c^2(1-p) & -c\sqrt{1 - c^2}(1-p) & 0 \\
0 & -c\sqrt{1 - c^2}(1-p) & (1 - c^2)(1-p) & 0 \\
0 & 0 & 0 & 0
\end{pmatrix}
\end{equation}
And the correlation matrix will be:
\begin{equation}
K = \begin{pmatrix}
-2c\sqrt{1-c^2}(1-p) & 0 & 0 \\
0 & -2c\sqrt{1-c^2}(1-p) & 0 \\
0 & 0 & 1-2p
\end{pmatrix}
\end{equation}
The beta parameter is the same as in the phase damping case, but the alpha parameter will be given by $\alpha = |1-2p|$. We take the modulus in order to ensure the positivity of the singular values. The function $f(\phi)$ will be given this time by:
\begin{equation}
    f(\phi)= \frac{\beta^2}{\alpha^2}=4c^2(1-c^2)(1-p)^2(|1-2p|)^{-2}
\end{equation}
The curve describing the effects of amplitude damping is presented below:
\begin{figure}[H]
    \centering
    \includegraphics[scale=0.50]{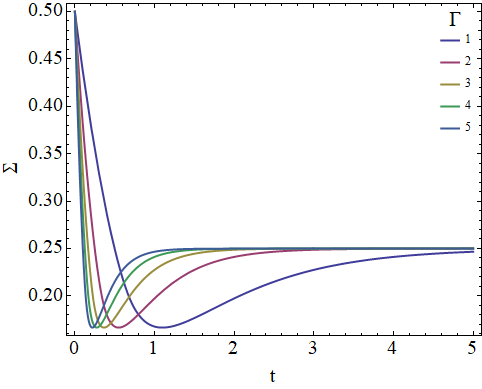}
    \caption{Decay of average correlation over time due to amplitude damping for various decoherence rates. The curve crosses the $\Sigma = 1/4$ threshold, indicating non-classical states.}
    \label{fig:enter-label}
\end{figure}
The graph illustrates the effects of energy dissipation on the time evolution of the average correlation. We note that initially all curves start with $\Sigma$ values above the nonclassicality threshold of $1/4$, indicating that the system is in a nonclassical state. As time progresses, the various rates of amplitude damping cause $\Sigma$ to decrease, demonstrating the loss of quantum correlations. Unlike phase damping, amplitude damping encompasses energy dissipation, which can lead to a complete loss of nonclassicality, as evidenced by $\Sigma$ potentially dropping below the $1/4$ threshold.

The decay trajectories of $\Sigma$ under different amplitude damping rates show that higher values of $\Gamma$ correspond to a more rapid decline in nonclassicality. This is a quantifiable demonstration of how the system's interaction with its environment accelerates the transition from quantum to classical states. The distinct rates of decay not only reflect the sensitivity of nonclassical states to environmental interactions but also highlight the temporal aspect of quantum coherence. We also point out that the effect of amplitude damping is more pronounced than that of phase damping presented earlier. This quantum channel involves energy dissipation which can significantly impact the coherence properties of quantum systems\cite{ampdamp1,ampdamp2,ampdamp3} more than phase damping does. In the study \cite{ampdamp-jaynes} for example, the authors explored the dynamics of a Jaynes-Cummings model under phase-damping conditions and concluded that, while phase damping does lead to decoherence, the absence of energy exchange between the system and its environment results in a comparatively slower progression towards classical states. 

It is evident from Fig(2) that some states may retain their nonclassical character longer than others, dependent on the magnitude of $\Gamma$, which is crucial for quantum computing and communication tasks that rely on nonclassical states. The long-term behavior of the system, as depicted by the convergence of the curves towards a lower bound of $\Sigma$, suggests a stabilization of the average correlation in the presence of amplitude damping. This equilibrium state, potentially indicative of a residual quantum correlation or a complete transition to classicality, emphasizes the nuanced nature of quantum decoherence. Understanding the duration for which nonclassicality is preserved is of paramount importance for the practical implementation of quantum technologies, as it dictates the window of opportunity for harnessing quantum mechanical advantages.

\section{Conclusion}
In this work, we systematically investigated the impact of decoherence mechanisms, specifically phase and amplitude damping, on the average correlation of quantum states. Our analysis revealed distinct effects of these decoherence processes on quantum coherence and correlation, highlighting the differential impact of phase and amplitude damping. This distinction is crucial for understanding the resilience and vulnerability of quantum states under various decoherence scenarios, thereby informing strategies for preserving quantum information integrity in quantum computing and communication technologies. Furthermore, our findings underscore the significance of coherence and correlation as fundamental resources in quantum technology applications, suggesting that maintaining these properties is essential for the practical realization of quantum computational and informational tasks. The exploration of decoherence effects presented in this paper contributes to the broader field of quantum information science by providing a clearer understanding of how quantum systems interact with their environment and the implications for quantum technology development. Effects of phase damping and amplitude damping have been widely studied in the literature \cite{effect1,effect2,effect3} and future research in this area may focus on developing advanced quantum error correction techniques and decoherence-resistant designs to mitigate the adverse effects observed, thereby enhancing the performance and reliability of quantum-based technologies.

Furthermore, more work in this area can expand upon the foundational insights garnered from this study by exploring generalizations of phase and amplitude damping channels \cite{further1,further2}. This direction could yield a deeper understanding of how quantum systems interact with complex environments over time. Additionally, examining a broader array of noisy channels, such as depolarizing, dephasing, and correlated noise channels, could provide comprehensive insights into the resilience of quantum states and the efficacy of quantum information processing under diverse environmental interactions.

Exploring the interplay between different types of noise and quantum entanglement, coherence, and other quantum correlations within multipartite systems could uncover new strategies for preserving quantum information. This includes the potential discovery of quantum states that are naturally resistant to specific noise types or the development of dynamic control techniques to adaptively protect quantum information.

\nocite{*}
\bibliographystyle{unsrt} 
\bibliography{main} 

\end{document}